\documentclass[showkeys,showpacs,superscriptaddress]{revtex4}
\usepackage{amsmath}
\usepackage{amssymb}
\usepackage{bm}
\usepackage{graphicx}
\allowdisplaybreaks

\begin{document}

\title{
Thermodynamics and statistical physics of quasiparticles within the quark-gluon plasma model
}

\author{
Vladimir Dzhunushaliev
}
\email{v.dzhunushaliev@gmail.com}
\affiliation{
Department of Theoretical and Nuclear Physics,  Al-Farabi Kazakh National University, Almaty 050040, Kazakhstan
}
\affiliation{
Institute of Experimental and Theoretical Physics,  Al-Farabi Kazakh National University, Almaty 050040, Kazakhstan
}
\affiliation{
Academician J.~Jeenbaev Institute of Physics of the NAS of the Kyrgyz Republic, 265 a, Chui Street, Bishkek 720071, Kyrgyzstan
}
\affiliation{
Institut f\"ur Physik, Universit\"at Oldenburg, Postfach 2503
D-26111 Oldenburg, Germany
}

\author{Vladimir Folomeev}
\email{vfolomeev@mail.ru}
\affiliation{
Institute of Experimental and Theoretical Physics,  Al-Farabi Kazakh National University, Almaty 050040, Kazakhstan
}
\affiliation{
Academician J.~Jeenbaev Institute of Physics of the NAS of the Kyrgyz Republic, 265 a, Chui Street, Bishkek 720071, Kyrgyzstan
}
\affiliation{
Institut f\"ur Physik, Universit\"at Oldenburg, Postfach 2503
D-26111 Oldenburg, Germany
}

\author{
Tlekkabul Ramazanov
}
\affiliation{
Institute of Experimental and Theoretical Physics,  Al-Farabi Kazakh National University, Almaty 050040, Kazakhstan
}

\author{
Tolegen Kozhamkulov
}
\affiliation{
Institute of Experimental and Theoretical Physics,  Al-Farabi Kazakh National University, Almaty 050040, Kazakhstan
}

\date{\today}

\begin{abstract}
We consider thermodynamic properties of a quark-gluon plasma related to quasiparticles having the internal structure. For this purpose, we employ a possible analogy between quantum chromodynamics and non-Abelian Proca-Dirac-Higgs
theory. The influence of characteristic sizes of the quasiparticles on such thermodynamic properties of the quark-gluon plasma like
the internal energy and pressure is studied. Sizes of the quasiparticles are taken into account in the spirit of the van der Waals equation
but we take into consideration that the quasiparticles have different sizes, and the average value of these sizes depends on temperature.
It is shown that this results in a change in the internal energy and pressure of the quark-gluon plasma. Also, we show that, when the temperature increases,
the average value of characteristic sizes of the quasiparticles increases as well. This leads to the occurrence of a phase transition
at the temperature at which the volume occupied by the quasiparticles is compared with the volume occupied by the plasma.
\end{abstract}



\date{\today}

\maketitle

\section{Introduction}

Quark-gluon plasma (QGP) can be defined as a high-temperature state of a non-Abelian SU(3)
gauge field and quarks where hadrons are broken apart. The transition from a hadronic state to the QGP takes place at some
critical temperature $T_c$. A full theoretical treatment of the QGP is still absent, and one might expect that such a description
can be done only within the framework of the nonperturbative quantization of such strongly nonlinear system like quantum chromodynamics (QCD).
For this reason, every approximate description clarifying any characteristic features of the QGP may be helpful.

In the present paper, we study the role of quasiparticles in the QGP. Lattice~\cite{Karsch:2002wv,Karsch:2000ps,Laursen:1987eb,Koma:2003hv,Bornyakov:2003vx}
and analytical investigations~\cite{Shuryak:2004tx,Liao:2006ry,Ramamurti:2017fdn,Ramamurti:2018evz,Shuryak:2018ytg}
indicate that the QGP contains various quasiparticles: monopoles, dyons, binary bound states [quark-quark ($qq$), quark-antiquark ($q \bar q$),
gluon-gluon ($gg$), quark-gluon ($qg$), etc.]. Quasiparticles are extended objects having their own energy and sizes. We wish to show here that:
(a)~their own energy causes a change in the internal energy of the QGP (within the approximation used in the present paper, the QGP is modeled as a gas consisting of quasiparticles);
(b)~nonzero sizes cause a change in the equation of state.

The item (b) can be easily explained: the volume where particles are in motion differs from the geometric volume $V$ where the QGP  is located. This is entirely similar to that happens in real gases described by the van der Waals equation. In analogy to real gases, one can introduce a correction to the volume as follows: $V \rightarrow (V - V_q)$, where $V_q$ is the volume occupied by all quasiparticles.  The significant difference from real gases is the fact that in our case $V_q$ becomes a function of temperature; this happens because when the temperature is changed, the internal structure (i.e., the distribution of fields creating a quasiparticle, as well as the energy and characteristic sizes) of the quasiparticles is also changed. It is evident that, within such an approach, the critical temperature $T_c$ occurs as a temperature at which the quasiparticles break apart  and a transition to a state without quasiparticles takes place.

In Ref.~\cite{Dzhunushaliev:2019ham}, arguments in favour of that QCD can exhibit general similarities to non-Abelian Proca-Dirac-Higgs theory are given. In particular, it was shown there that within non-Abelian Proca-Dirac-Higgs theory there is a mass gap, at least for one frozen parameter. Since QCD should possess a mass gap, it is assumed that the mechanism of creation of such a gap in QCD may be the same. In non-Abelian Proca-Dirac-Higgs theory, the presence of the mass gap is related to the nonlinear Dirac equation. It is therefore assumed in Ref.~\cite{Dzhunushaliev:2019ham} that in QCD the nonlinear Dirac equation occurs as a result of approximate description of the interaction between see quarks and gluons. This can happen as follows. In the Lagrangian, the interaction between quarks and gluons is described by the term
$
	\hat{\bar \psi} \lambda^B \hat{A}_\mu \hat{\psi}
$, where
$
\hat{\psi} = \left\langle \hat{\psi} \right\rangle + \widehat{\delta \psi},
\hat A_\mu = \left\langle \hat{A}_\mu \right\rangle + \widehat{\delta A}_\mu
$. Here
$
	\left\langle \hat{\psi} \right\rangle$ and
	$\left\langle \hat{A}_\mu \right\rangle
$ are valence quarks and gluons, and
$
 \widehat{\delta \psi}$ and $\widehat{\delta A}_\mu
$ are sea quarks and gluons; $\lambda^B $ is the Gell-Mann matrices.
One can assume that the quantum average of the term
$
	\left\langle
		\widehat{\delta \bar{\psi}} \lambda^B \widehat{\delta A}_\mu
		\widehat{\delta \psi}
	\right\rangle
$
will \emph{approximately} look like
$
	\left\langle
		\widehat{\delta \bar{\psi}} \lambda^B \widehat{\delta A}_\mu
		\widehat{\delta \psi}
	\right\rangle \approx
	\phi \left(
		\bar \xi \xi
	\right)^2
$, where the scalar field $\phi$  \emph{approximately} describes the sea gluons and the spinor field $\xi$ \emph{approximately} describes the sea quarks. Thus, in QCD, there can occur the nonlinear Dirac equation which \emph{approximately} describes the interaction between sea quarks and gluons.

Note also that, in the van der Waals gas, particles are assumed to be hard spheres which are scattered elastically with one another
because of the presence of some repulsive forces. In our case, we consider a QGP consisting of quasiparticles created  by sea quarks.
Such quasiparticles are a source of gauge (Proca) fields and, as demonstrated in
Ref.~\cite{Dzhunushaliev:2019ham}, they are magnetic monopoles. For the sake of simplicity, in the present paper
we consider a plasma consisting of quasiparticles which are modeled with
no account taken of gauge fields created by the quasiparticles.
But for understanding the nature of the repulsive forces between quasiparticles, it is necessary to realize that the quasiparticles must create a color magnetic field
whose presence will result in the appearance of the  repulsive forces between them.
Also, notice that, for simplicity, we consider the plasma consisting of quasiparticles (monopoles) having magnetic charges of the same sign.

In order to study thermodynamic properties of the QGP, we employ here this analogy between the QGP in QCD and a plasma in non-Abelian Proca-Dirac-Higgs theory.
In such theory, the simplest particlelike solution is the solution containing only a spinor field, presumably describing a virtual pair of quarks~-- a spinball. In the QGP, such an object is a quasiparticle.

\section{ Control parameters
}

In order to describe the properties of the QGP consisting of quasiparticles, one can use the control parameters
known from the theory of strongly correlated systems: the coupling parameter $\Gamma$, the degeneracy parameter $\chi$, and the Brueckner parameter $r_s$~\cite{Bonitz}.
As applied to the QGP, they are
\begin{eqnarray}
	\Gamma &=& \frac{\overline{W_q}}{k T} ,
\label{4_10}\\
	\chi &=& n_q \overline{v_q} ,
\label{4_20}\\
	r_s &=& \frac{\overline{l_s}}{\overline{l_q}}.
\label{4_30}
\end{eqnarray}
Here $\overline{W_q}$ is the average value of the quasiparticle energy; $n_q$ is the concentration of quasiparticles;
$\overline{v_q}$ is the average value of the volume of one quasiparticle; $\overline{l_s}$ is the average distance between the quasiparticles;
$\overline{l_q}$ is the characteristic size of quasiparticles. It is evident that  $\overline{l_q} \approx \overline{v_q}^{1/3}$.

\section{Statistical integral}

We consider a QGP model containing, apart from valence quarks and gluons, quasiparticles having finite sizes and the internal energy
which is created by the field from which a quasiparticle is constructed. For example, this can be
quasiparticles mentioned in Ref.~\cite{Shuryak:2004tx}: monopoles, dyons, binary bound states [quark-quark ($qq$), quark-antiquark ($q \bar q$),
gluon-gluon ($gg$), quark-gluon ($qg$), etc.]. Apparently, one can construct such quasiparticles only using nonpertubative QCD.
Here we will employ an assumption stated in Ref.~\cite{Dzhunushaliev:2019ham},
according to which non-Abelian Proca theory containing nonlinear spinor and Higgs fields possesses some properties similar to QCD. It was shown in Ref.~\cite{Dzhunushaliev:2019ham} that in such a theory there is a mass gap,
at least for one fixed parameter determining particlelike solutions within this theory. The presence of the mass gap is related to the nonlinear
Dirac equation, as it was
discovered in
Refs.~\cite{Finkelstein:1951zz,Finkelstein:1956}. It was supposed in Ref.~\cite{Dzhunushaliev:2019ham} that the mass gap in QCD has the same nature,
and the nonlinear Dirac equation occurs as a result of approximate description of the interaction between sea quarks and gluons.
This means that non-Abelian Proca theory plus the nonlinear Dirac equation plus the Higgs field and nonperturbative QCD may have common properties.

We therefore consider here a plasma consisting of valence quarks and Proca gluons and containing also quasiparticles described by particlelike solutions from non-Abelian Proca-Dirac-Higgs theory.
Our purpose is to show that the presence of the internal structure of the quasiparticles leads to a considerable change in such thermodynamic quantities like the internal energy and pressure.

The total energy of a nonrelativistic particle $W$ consists of two parts: the energy $W_q$ (associated with the energy
related to the presence of the internal field structure) and the kinetic energy $W_k$. That is,
\begin{equation}
	W =W_q + W_k\equiv W_q + \frac{c^2 p^2}{2 W_q} ,
\label{1_10}
\end{equation}
where we have taken into account that the mass of a quasiparticle is $m_q = W_q/c^2$.

For simplicity, we will consider quasiparticles assuming that they do not interact with each other and with valence quarks and gluons containing in the QGP.
The finiteness of quasiparticle sizes must be also taken into account.
To do this, to a first approximation, we use the idea coming from the van der Waals equation: particles move not in the volume $V$ but in a smaller volume
$\left(V - V_q\right)$, where $V_q$ is the volume occupied by the quasiparticles. It is evident that  the volume occupied by $N$  quasiparticles is
\begin{equation}
	 V_q = \sum_{i=1}^N v_i ,
\label{1_40}
\end{equation}
where $v_i$ is the characteristic volume occupied by the $i$-th quasiparticle.

Integration over coordinates in the statistical integral, by taking into account the sum \eqref{1_40}, runs into great difficulty.
Therefore, to simplify the problem, we estimate the sum  \eqref{1_40} as follows:
\begin{equation}
	 V_q \approx N \overline{v_q} =
	 V n_q \overline{v_q} = V \chi ,
\label{1_50}
\end{equation}
where we have taken into account the definition  \eqref{4_20} for the degeneracy parameter $\chi$.
This simplification permits us to write the statistical integral as a product of the corresponding integrals for every single particle,
\begin{equation}
	Z(T) \approx Z_{(\text{quarks + gluons)}} \left[
		\int dV dp_x dp_y dp_z \rho(\bm{\gamma})d  \bm{\gamma}\,
		e^{ - \frac{W_q(\bm{\gamma}) + \frac{c^2 p^2}{2 W_q(\bm{\gamma})}
		- \Delta}{k T}}
	\right]^N = Z_{(\text{quarks + gluons})}
	Z_{\text{quasiparticles}}.
\label{1_20}
\end{equation}
Here $Z_{\text{(quarks + gluons)}}$ is the statistical integral for valence quarks and gluons; $Z_{\text{quasiparticles}}$ is the statistical integral associated with the presence of the internal structure of quasiparticles;
$\bm \gamma$  is the set of  parameters on which the energy of quasiparticles depends; $T$ is the temperature; $\rho(\bm{\gamma})$  is the density of states.
The constant $\Delta$ corresponds to the minimum energy of a quasiparticle (the mass gap) from which the energy will be reckoned (see below).

For simplicity, in the present paper, we consider only the case of noninteracting quasiparticles. Our main purpose is to study the effects related
to the presence of particles having the internal structure: the internal energy and finite sizes. The presence of the internal energy is  taken into account
by the term $W_q$ in Eq.~\eqref{1_10}. Then, using the approximation \eqref{1_50}, the integration over the volume in \eqref{1_20} gives us $(V - V_q)$,
and the statistical integral takes the  form
\begin{eqnarray}
	Z_{\text{quasiparticles}}(T) &=& Z_0 T^{3 N/2}
	\left( V - V_q \right)^N \left[
		\int W_q^{3/2}(\bm{\gamma})\rho\left(\bm{\gamma}\right)
		e^{- \frac{W_q(\bm{\gamma}) - \Delta}{ k T}} d  \bm{\gamma}
	\right]^N
\nonumber\\
	&=& Z_0 T^{3 N/2} \left(
		V - V_q
	\right)^N \left( Z_q \right)^N= Z_0 T^{3 N/2} V^N
	\left(1 - \chi \right)^N \left( Z_q \right)^N .
\label{1_30}
\end{eqnarray}
Here  $Z_q$ is the statistical integral per one quasiparticle and all dimensional constants are collected in the normalization constant $Z_0$.
For simplicity, in performing numerical calculations in subsequent sections,
we suppose that all properties of quasiparticles depend only on one parameter  $\gamma$ (i.e., we will consider the case where the set of parameters  $\bm{\gamma}$ contains only one parameter).
It is evident that characteristic sizes of the quasiparticles (and hence the volume) depend on temperature because when the temperature changes,
the average volume will also change:
$
	 \overline{v_q} = \overline{v_q}(T)
$.

The internal energy of the quasiparticles is defined as follows:
\begin{equation}
	U_\text{quasiparticles} =
	\frac{1
	}{Z_{\text{quasiparticles}}}
	\int \sum \limits_{i=1}^N\left[
			W_{q,i}\left( \bm{\gamma}_i \right) + W_{k,i}-\Delta
		\right]
		e^{- \frac{\sum \limits_{i=1}^N\left[
			W_{q,i}\left( \bm{\gamma}_i \right) + W_{k,i}-\Delta
		\right]}{kT}}\prod \limits_{i=1}^N d V_i d \vec{p}_i \rho(\bm{\gamma}_i)d \bm{\gamma}_i ,
	\label{1_33}
\end{equation}
where $W_{q,i}$ and $W_{k,i}$ correspond to energies of the $i$-th particle.
Since
\begin{equation}
	\int e^{ - \frac{
		W_q\left( \gamma_i \right) + W_k 		
		}{k T}} dV dp_x dp_y dp_z d \bm{\gamma} \sim
	T^{3 /2} \left(
		V - V_q
	\right)
	\int W_q^{3/2}\left( \bm{\gamma} \right)
	e^{- \frac{W_q\left( \bm{\gamma} \right)}{k T}} d \bm{\gamma},
\label{1_35}
\end{equation}
the expression for the internal energy \eqref{1_33} takes the following form:
\begin{equation}
\begin{split}
	\frac{U_\text{quasiparticles}}{N} \equiv U_\text{quasiparticle} = &
	\frac{
		\int W_q^{5/2}\left( \bm{\gamma} \right)
		e^{- \frac{W_q\left( \bm{\gamma} \right)}{k T}} \rho(\bm{\gamma})d \bm{\gamma} 		
	}
	{
		\int W_q^{3/2}\left( \bm{\gamma} \right)
		e^{- \frac{W_q\left( \bm{\gamma} \right)}{k T}} \rho(\bm{\gamma})d \bm{\gamma}
	} -\Delta+ \frac{c^2}{2} \frac{
		\int \frac{p^2}{W_q\left( \bm{\gamma} \right)}
		e^{
			- \frac{W_q\left( \bm{\gamma} \right) + W_k}{kT}
		}\rho(\bm{\gamma}) dp_x dp_y dp_z d \bm{\gamma}
	}
	{
		\int e^{- \frac{W_q\left( \bm{\gamma} \right) + W_k}{kT}}
		\rho(\bm{\gamma})dp_x dp_y dp_z d \bm{\gamma}
	}
\\
	&
	=\frac{
		\int W_q^{5/2}\left( \bm{\gamma} \right)
		e^{- \frac{W_q\left( \bm{\gamma} \right)}{k T}} \rho(\bm{\gamma})d \bm{\gamma} 		
	}
	{
		\int W_q^{3/2}\left( \bm{\gamma} \right)
		e^{- \frac{W_q\left( \bm{\gamma} \right)}{k T}} \rho(\bm{\gamma})d \bm{\gamma}
	} -\Delta+ \frac{3}{2} kT.
\end{split}
\label{1_37}
\end{equation}
The gas pressure of one quasiparticle is defined as follows:
\begin{equation}
	\frac{p_\text{quasiparticles}}{N} = - \frac{1}{N}\frac{\partial F_{\text{quasiparticles}}}{\partial V} =
	\frac{k T}{ V - V_q } =
	\frac{k T}{ V \left( 1 - \chi \right) } =
	\frac{k T}{ V \left( 1 -  n_q \overline{v_q} \right) }.
\label{3_15}
\end{equation}

We emphasize that since the average volume of one quasiparticle depends on temperature, the equation of state \eqref{3_15} will differ strongly from the van der Waals equation.
Also note that below we choose $\rho(\bm{\gamma})=\text{const}$.

\section{A qualitative estimate  of the statistical integral $Z_q$ in the presence of the mass gap
}

In this section we estimate the statistical integral $Z_q$ and quantities related to this integral. To do this, we assume that in the region
where the parameters $\bm{\gamma}$ are changed, the energy $W_q$ tends to infinity on the boundary of this volume,
$
	\left( W_q\right)_{\partial V_q} \rightarrow \infty
$, and the dependence of the energy on the parameters
$W_q\left( \bm{\gamma} \right)$ has an inverted bell shape. This means that inside the volume $V_q$ there is a point where the energy takes its minimum value $\Delta$.
This also means that there is some characteristic volume $\Omega_0$ outside which the energy $W_q$ goes to zero sufficiently fast.
If the minimum value of the energy is positive,  i.e., $\Delta > 0$, one can say about the presence of a mass gap for such quasiparticles. Then one may estimate the statistical integral per one particle as follows:
\begin{equation}
	Z_{\text{quasiparticle}} \approx Z_0 T^{3/2}
	\left( V - V_q \right)
	 W_q^{3/2}\left( \Delta \right)
	e^{- \frac{W_q\left( \Delta \right) }{T}} \Omega_0(T) .
\label{5_10}
\end{equation}
Here, we have taken into account that the size of the volume $\Omega_0$ in the parameter space  $\bm\gamma$ depends on temperature:
$\Omega_0 = \Omega_0(T)$. Substituting the expression \eqref{5_10} in Eqs.~\eqref{1_37} and \eqref{3_15}, we have
\begin{eqnarray}
	U_{\text{quasiparticle}} &\approx& \frac{3}{2} kT +
	W_q\left( \Delta \right) ,
\label{5_20}\\
	p_{\text{quasiparticle}} &=& \frac{k T}{ V - V_q } =
	\frac{k T}{ V \left( 1 - \chi \right) } =
	\frac{k T}{ V \left( 1 -  n_q \overline{v_q} \right) }.
\label{5_30}
\end{eqnarray}

\section{Fermion-Proca gluon plasma: the simplest model}

In the above discussion, we have argued that the fermion-Proca gluon plasma may have common features with the QGP in QCD.
Consistent with this, in this section we give a more detailed study of our model for such plasma.

The construction of statistical physics for a fermion-Proca gluon plasma is an extremely complicated problem since the corresponding particlelike
solutions depend on a number of parameters. At the moment, it appears to be impossible to find the energy spectrum of these solutions
for the whole set of the parameters. We therefore consider a simplified model for such a problem.
To do this, we freeze the degrees of freedom related to a Proca field and consider only a spinor field described by the nonlinear Dirac equation.
Our purpose will be to calculate the statistical integral for a gas consisting of spinballs: a fermion-Proca gluon plasma where the Proca field is frozen.

Since we freeze the degrees of freedom of the Proca field, a quasiparticle in the fermion-Proca gluon plasma is described by the nonlinear
Dirac equation \eqref{2_10} (recall that, according to Ref.~\cite{Dzhunushaliev:2019ham}, it is assumed that this equation approximately describes the
interaction between see quarks and gluons) without the Proca Field $A^a_\mu$ (cf. the equations of Ref.~\cite{Dzhunushaliev:2019ham}
where $A^a_\mu=0$ and the scalar field is assumed to be constant),
\begin{equation}
	i \hbar \gamma^\mu
		\partial_\mu  \psi + \Lambda \psi
	\left(
		\bar \psi \psi
	\right) - m_f c \psi = 0 .
\label{2_10}
\end{equation}
where $m_f$ is the mass of the spinor field and $\Lambda$ is a constant.
It must be mentioned here that the approach with the use of Eq.~\eqref{2_10} is formally similar to
the Nambu-Jona-Lasinio (NJL) model involving the corresponding equation. The principle difference of our approach from
the NJL model is that (a) our Eq.~\eqref{2_10} describes virtual (sea) quarks, whereas
the NJL model is an effective field theory of nucleons and mesons, and  (b) in Eq.~\eqref{2_10}, there is a massive term,
which is absent in the NJL equation.

Introducing
$
	l_0 = \frac{\hbar}{m_f c}
$ and redefining
$
	\psi \sqrt{\frac{\Lambda}{m_f c}} \rightarrow \psi
$,
the Dirac equation  \eqref{2_10} takes the following dimensionless form:
\begin{equation}
	\left[
		i \gamma^\mu \partial_\mu \psi +
		\left(
			\bar \psi \psi
		\right) - 1
	\right] \psi = 0.
\label{2_20}
\end{equation}
\textit{Ansatz} for the doublet of the spinor field is taken in the form~\cite{Dzhunushaliev:2019ham}
 \begin{equation}
	\psi^T = \frac{e^{-i \frac{ E t}{\hbar}}}
	{r \sqrt{2}} \begin{Bmatrix}
		\begin{pmatrix}
			0 \\ - u \\
		\end{pmatrix},
		\begin{pmatrix}
			u \\ 0 \\
		\end{pmatrix},
		\begin{pmatrix}
			i  v \sin \theta e^{- i \varphi} \\
			- i v \cos \theta \\
		\end{pmatrix},
		\begin{pmatrix}
			- i v \cos \theta \\
			 - i v \sin \theta e^{i \varphi} \\
		\end{pmatrix}
	\end{Bmatrix},
\label{2_30}
\end{equation}
where $ E/\hbar$ is the spinor frequency. Substitution of \eqref{2_30} in \eqref{2_20} yields the following equations describing a spinball as a quasiparticle in the spinball plasma:
\begin{eqnarray}
	\tilde v' + \frac{\tilde v}{\bar x} &=& \tilde u \left(
		- 1 + \bar E + \frac{\tilde u^2 - \tilde v^2}{\bar x^2}
	\right) ,
\label{2_40}\\
	\tilde u' - \frac{\tilde u}{\bar x} &=& \tilde v \left(
		- 1 -\bar  E + \frac{\tilde u^2 - \tilde v^2}{\bar x^2}
	\right),
\label{2_50}
\end{eqnarray}
where we have introduced the following dimensionless quantities:
$
	\bar x=r/l_0, \bar{E}= (l_0/\hbar c)E,
	\left( \tilde u, \tilde v \right) = \sqrt{l_0} \left( u,v \right)
$.

\begin{figure}[t]
\begin{minipage}[t]{.45\linewidth}
	\begin{center}
		\includegraphics[width=1\linewidth]{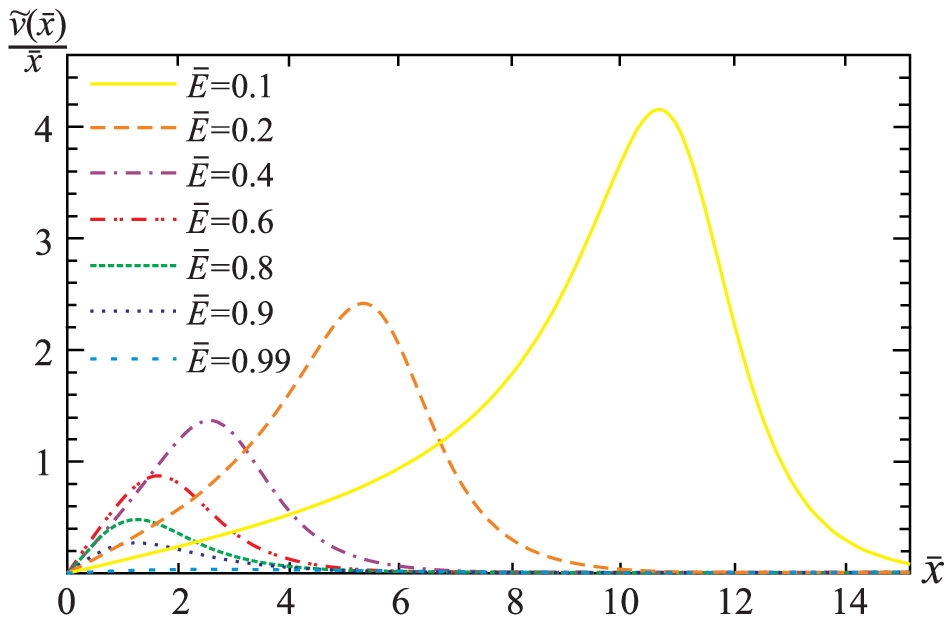}
	\end{center}
\vspace{-0.5cm}		
\caption{
	Family of the spinball solutions $\tilde v(\bar x)$ for different values of the parameter $\bar E$.
}
\label{v_x_family}
\end{minipage}\hfill
\begin{minipage}[t]{.45\linewidth}
	\begin{center}
		\includegraphics[width=1\linewidth]{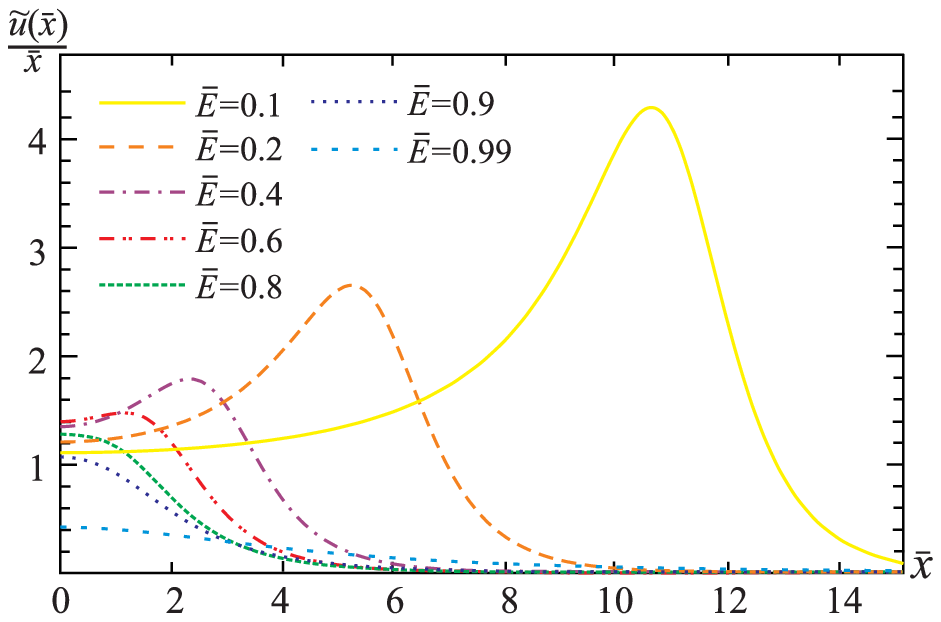}
	\end{center}
\vspace{-0.5cm}
\caption{	
	Family of the spinball solutions $\tilde u(\bar x)$ for different values of the parameter $\bar E$.
}
\label{u_x_family}
\end{minipage}\hfill
\begin{minipage}[t]{.45\linewidth}
	\begin{center}
		\includegraphics[width=1\linewidth]{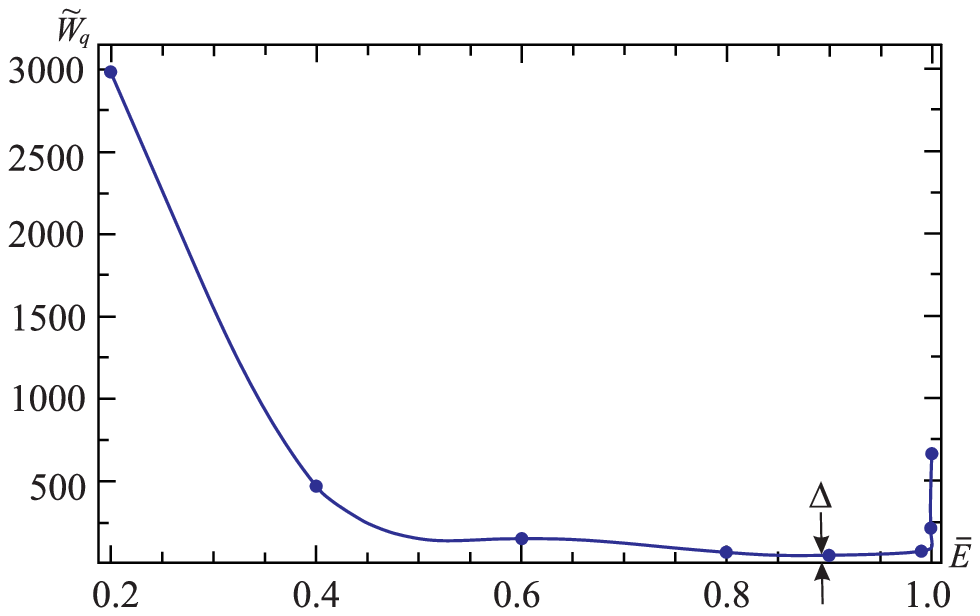}
	\end{center}
\vspace{-0.5cm}		
\caption{
	The dependence of the spinball energy	$\tilde W_q$ from \eqref{2_70} on $\bar E$ (for $\tilde \Lambda=1$)
and the location of the  mass gap~$\Delta$.
}
\label{massgapSB}
\end{minipage}\hfill
\begin{minipage}[t]{.45\linewidth}
	\begin{center}
		\includegraphics[width=1\linewidth]{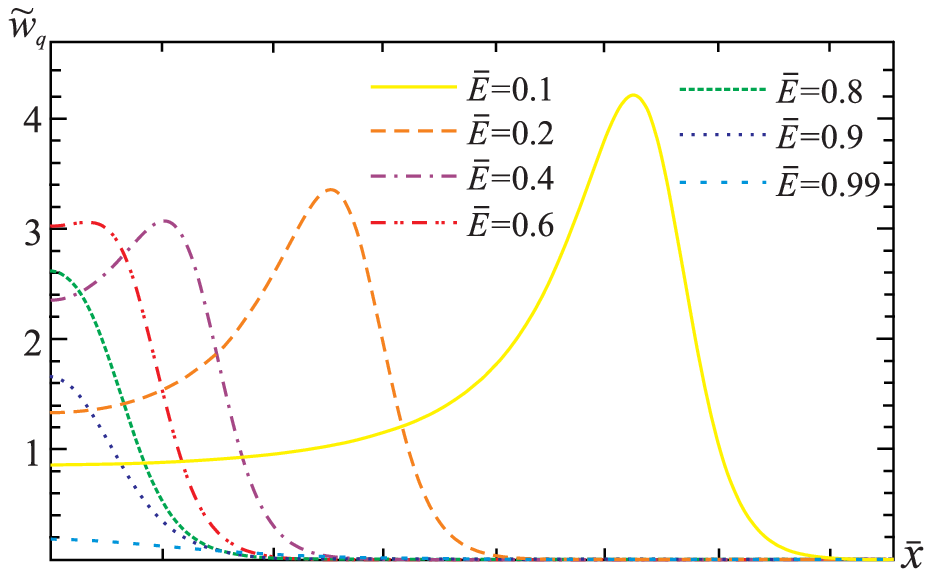}
	\end{center}
\vspace{-0.5cm}		
\caption{	
	Family of the spinball energy densities $\tilde w_q(\bar x)$ from \eqref{2_60} for different values of the parameter $\bar E$.
}
\label{EnDenSB_QM}
\end{minipage}
\begin{minipage}[t]{.45\linewidth}
\begin{center}
	\includegraphics[width=1\linewidth]{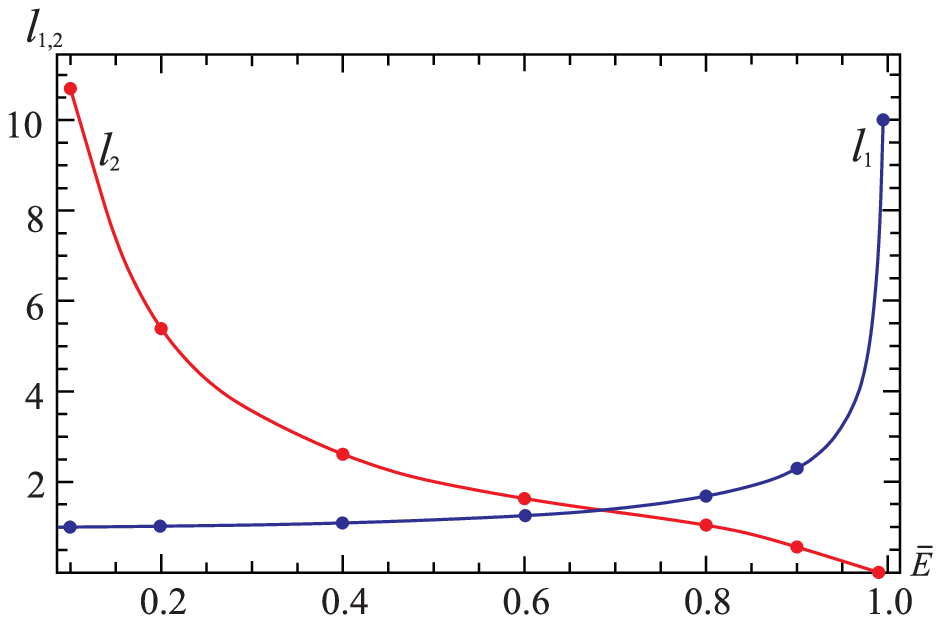}
\end{center}
\vspace{-0.5cm}	
\caption{The dependence of the characteristic sizes $l_{1,2}$ of the spinball for the left- and right-hand branches of the energy spectrum shown in Fig.~\ref{massgapSB}.
}
\label{l_1_2}
\end{minipage}
\end{figure}

Solving Eqs.~\eqref{2_40} and \eqref{2_50} for different values of the parameter $\bar E$, we get the data given in Table~\ref{mgsp}.
Figs.~\ref{v_x_family} and \ref{u_x_family} show the families of the corresponding solutions for the functions $\tilde v(\bar x)$ and $\tilde u(\bar x)$.
The profiles of the dimensionless energy density are shown in Fig.~\ref{EnDenSB_QM}. (Notice here that Figs.~\ref{v_x_family}-\ref{l_1_2} are taken from Ref.~\cite{Dzhunushaliev:2018zqe}.)

\begin{table}[h]
\scalebox{1.}{
\begin{tabular}{|c|c|c|c|c|c|c|c|c|c|}
	\hline
	\rule[-1ex]{0pt}{2.5ex}
	$\bar E$&0.1&0.2&0.4&0.6&0.8&0.9&0.99&0.999&0.9999\\
	\hline
	\rule[-1ex]{0pt}{2.5ex}
	$\tilde u_1$&1.103640998&1.20242169&1.34371184&1.389621&1.2745556&1.06477&
		0.419164&0.1366701&0.0433583\\
	\hline
	\rule[-1ex]{0pt}{2.5ex}
	$\tilde W_q$&18585&2985.04&470.563&152.677&66.4478&49.285&
	74.2536&213.603&668.854\\
	\hline
	\rule[-1ex]{0pt}{2.5ex}
	$l_2$&10.7004&5.39175&2.61383&1.6336&1.04804&0.563584&
	0.000946652&0&0\\
	\hline
\end{tabular}
}
\caption{
	Eigenvalues $\tilde u_1$ (the expansion coefficient in the vicinity of the center~\cite{Dzhunushaliev:2018zqe}) and the energy $\tilde W_q$ as functions of the parameter $\bar E$.
}
\label{mgsp}
\end{table}

Asymptotic behavior of the solutions shown in  Figs.~\ref{v_x_family} and \ref{u_x_family} is as follows:
\begin{eqnarray}
	\tilde u(\bar x) &\approx& \tilde u_\infty e^{- \bar x \sqrt{1 - \bar E^2}} ,
\label{2_54}\\
	\tilde v(\bar x) &\approx& \tilde v_\infty e^{- \bar x \sqrt{1 - \bar E^2}},
\label{2_58}
\end{eqnarray}
where $\tilde u_\infty, \tilde v_\infty$ are some constants. These expressions permit us to determine the characteristic size of such particlelike solutions (quasiparticles) as $\bar E \rightarrow 1$:
\begin{equation}
	l_1 \approx \frac{l_0}{\sqrt{1 - \bar E^2}} .
\label{2_59}
\end{equation}
It is seen that when $\bar E \rightarrow 1$ the characteristic size of the quasiparticle tends to infinity: $l_1 \rightarrow \infty$.
From the analysis of Figs.~\ref{v_x_family}, \ref{u_x_family}, and \ref{EnDenSB_QM},
it also follows that when $\bar E \rightarrow 0$
the characteristic size of the quasiparticle $l_2 \rightarrow \infty$. But if in the first case by the characteristic size $l_1$ we mean the semi-width of the curves
$\tilde u(\bar x)$ or $\tilde v(\bar x)$, in the second case by the characteristic size $l_2$ we mean the distance from the maximum of the curves $\tilde u(\bar x)$ or $\tilde v(\bar x)$
to the origin of coordinates. It is seen that the location of this maximum shifts to the right from the origin of coordinates as $\bar E\to 0$.
The dependencies of $l_{1}$ and $l_{2}$ on the energy $\bar E$ are shown in Fig.~\ref{l_1_2}. The numerical results  indicate that
when $\bar E \rightarrow 0, 1$ the characteristic sizes $l_{1, 2} \rightarrow \infty$.

The energy density of a quasiparticle in the spinball plasma is (cf. Ref.~\cite{Dzhunushaliev:2018zqe})
\begin{equation}
	w_q = \frac{m_f^4 c^5}{\hbar^3 {\tilde \Lambda}^{1/2}}
	\left[
		\bar E \frac{\tilde u^2 + \tilde v^2}{\bar x^2} +
		\frac{1}{2} \left(
			\frac{\tilde u^2 - \tilde v^2}{\bar x^2}
		\right)^2
	\right] = \frac{m_f^4 c^5}{\hbar^3 {\tilde \Lambda}^{1/2}}
	\tilde w_q,
\label{2_60}
\end{equation}
where the dimensionless $\tilde \Lambda=\Lambda/\left(m_f c\, l_0^3\right)$.
The energy of the spinball is
\begin{equation}
	W_q(E) = 4 \pi \frac{m_f c^2}{\sqrt{\tilde \Lambda}} \int_{0}^{\infty} x^2 \tilde w_q(\bar E) dx =
	4 \pi \frac{m_f c^2}{\sqrt{\tilde \Lambda}} \tilde W_q(\bar E).
\label{2_70}
\end{equation}
Here, we have introduced the dimensionless energy $\tilde W_q(\bar E)$ and emphasized that the spinball energy depends only on one parameter $\bar E$.

Summarizing, one can say that the quasiparticle (spinball) has the following properties:
\begin{itemize}
\item
	The energy spectrum of the spinball is restricted from below, i.e., this spectrum possesses a mass gap discovered in Refs.~\cite{Finkelstein:1951zz,Finkelstein:1956}.
\item
	The spinball energy depends on one parameter $\bar E \in [0, 1]$, and this energy tends to infinity as $\bar E \rightarrow 0, 1$.
\item
	The characteristic sizes of the spinball also tend to infinity as $\bar E \rightarrow 0, 1$.
\end{itemize}

It was supposed in Ref.~\cite{Dzhunushaliev:2019ham} that there is some relationship between non-Abelian Proca-Dirac-Higgs theory and QCD.
This relationship consists in the fact that in Proca-Dirac-Higgs theory there is a mass gap, at least for a fixed value of one of the parameters ensuring
the existence of particlelike solutions within such theory. The analysis indicates that the existence of the mass gap is related to the nonlinear Dirac equation:
there is no mass gap in the absence of spinor fields. In turn, numerical lattice calculations indicate that there is a mass gap in QCD; therefore, it is argued
in Ref.~\cite{Dzhunushaliev:2019ham} that perhaps the presence of the mass gap in QCD is related to the fact that the interaction between  sea quarks and
sea gluons can be approximately described using the nonlinear Dirac equation.
Also, this permits one to assume that a plasma in the
theory where quasiparticles are described by particlelike solutions found in  Ref.~\cite{Dzhunushaliev:2019ham} may serve as an approximation in describing the
properties of the QGP in QCD.

In subsequent sections, we will study changes in statistical physics and thermodynamics of plasma within
Proca-Dirac-Higgs theory induced only by spinballs.
This is because, a study of a gas consisting of more
complicated quasiparticles supported also by a non-Abelian Proca field is a much more difficult problem since the energy of such particlelike solutions
depends on a larger number of parameters; this greatly complicates a numerical study of statistical physics and thermodynamics of such a gas.

It is evident that the spinball gas approximation is a rough one: some statistical and thermodynamic characteristics of such a plasma
(Proca quark-gluon plasma) will be lost. In particular, this applies to the loss of a phase transition between hadronic matter and the Proca quark-gluon plasma.
This happens because a spinball contains only a spinor field but not a gluon field. Therefore, when the temperature drops down, hadronic matter does not
form.

In this case there is another phase transition: when the temperature rises, the size of a spinball  increases. For some temperature  $T_c$, this size becomes
comparable to the distance between quasiparticles. Then the typical distance between spinballs is defined as follows:
\begin{equation}
	l_s \sim \left( n_s \right)^{-1/3} .
\label{2_110}
\end{equation}
It is evident that in this case it is already impossible to describe our system as a plasma consisting of quasiparticles. This means that there is a phase transition
from a thermodynamic state  (spinball plasma) to a state where such quasiparticles disappear.

We can estimate a characteristic size of the spinball $l_q$ as
$\bar E \rightarrow 1$ using the formula~\eqref{2_59}. Unfortunately, there is no analytic formula to estimate the characteristic size
as $\bar E \rightarrow 0$. Therefore, in Sec.~\ref{toy_model_1}, we will discuss a toy model where the main characteristics of the spinball energy spectrum
 and the dependence \eqref{2_59} are kept and where we introduce an {\it ad hoc} dependence of $l_q$ on the parameter defining the particlelike
solution. Then, in Sec.~\ref{toy_model_2}, we will carry out similar calculations for a spinball plasma when the dependence $l_q(\bar E)$
is specified using the interpolation of the curves given in Fig.~\ref{l_1_2}.

\subsection{Toy model}
\label{toy_model_1}

To calculate the statistical integral \eqref{1_30}, let us assume that the energy of a quasiparticle is defined as follows:
\begin{equation}
	W_{q}(x) = \frac{\Delta}{1 - x^2}.
\label{6_a_10}
\end{equation}
Here $\Delta$ is the minimum value of the energy (the mass gap) and the parameter $x$ is the parameter $\gamma $
from \eqref{1_30}. As $x \rightarrow 1$ and $x \rightarrow - 1$, the characteristic size of such quasiparticles $l_q$
is defined as
\begin{equation}
	\frac{l_q}{l_0} = \frac{a}{\left( 1 - x \right)} +  \frac{b}{\left( 1 + x \right)} ,
\label{6_a_20}
\end{equation}
where $l_0$,  $a$, and $b$ are some constants. We have chosen the terms appearing in the right-hand side of Eq.~\eqref{6_a_20} so that they coincide qualitatively
with the dependencies $l_{1}$ and $l_2$ shown in Fig.~\ref{l_1_2}.
In particular, the characteristic sizes of a quasiparticle diverge for $x \rightarrow \pm 1$.

In numerical calculations of the internal energy $U$, of the pressure $p$, and of the average value of the characteristic volume $\overline{v_q}$
for one quasiparticle we use the formulas [see Eqs.~\eqref{1_37} and \eqref{3_15}]
\begin{eqnarray}
	\tilde U_{\text{quasiparticle}}(\tilde T)\equiv\frac{\tilde U_\text{quasiparticles}(\tilde T)}{N_q} &=& 	\frac{
			\int \tilde W_q^{5/2}\left(x \right)
			e^{- \frac{\tilde W_q\left( x\right)}{\tilde  T}} d x	
		}
		{
			\int \tilde W_q^{3/2}\left( x \right)
			e^{- \frac{\tilde W_q\left( x \right)}{\tilde  T}} d x
		} -\tilde \Delta+ \frac{3}{2} \tilde T ,
\label{6_a_30}\\
	\frac{\tilde p_\text{quasiparticles}(\tilde T)}{N_q} &=&
	\frac{\tilde T}{ \tilde V - \tilde V_q } =
	\frac{\tilde T}{ \tilde V \left( 1 - \chi \right) } =
	\frac{\tilde T}{ \tilde V \left( 1 -  n_q \overline{v_q} \right) },
\label{6_a_35}\\
	\frac{\overline{v_q}(\tilde T)}{v_0} &=& \frac{
			\int \left( \frac{l_q}{l_0} \right)^3
			\tilde{W}_q^{3/2}(x)
			e^{- \frac{\tilde W_q(x)}{\tilde T}} d x
		}{Z_q},
\label{6_a_40}
\end{eqnarray}
where we have introduced dimensionless $\tilde T=T/T_0$, $(\tilde W_q, \tilde \Delta, \tilde U)=(W_q, \Delta, U)/(k T_0)$,
$\tilde p_\text{quasiparticles}=\left[l_0^3/(k T_0)\right]p_\text{quasiparticles}$,
$\tilde V=V/l_0^3$.
Here $T_0$ is some characteristic temperature,
$v_0 = l_0^3$, and the partition function $Z_q$ is calculated according to the formula~\eqref{1_30},
\begin{equation}
	Z_q(\tilde T) = \int \tilde{W}_q^{3/2}(x)
		e^{- \frac{\tilde W_q(x)}{\tilde T}} d x .
\label{6_a_45}
\end{equation}

\begin{figure}[t]
\begin{minipage}[t]{.45\linewidth}
\centering
	\includegraphics[width=1\linewidth]{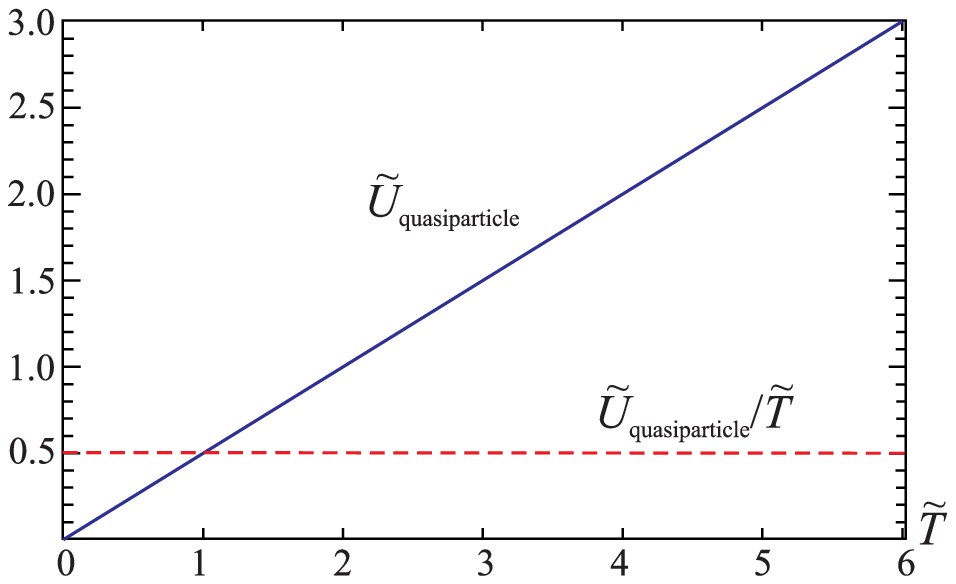}
\caption{
	The internal energy $\tilde U_{\text{quasiparticle}}$
(without allowance for the term corresponding to an ideal gas)
and the contribution $\tilde U_{\text{quasiparticle}}/\tilde T$ to the entropy	\eqref{entrop}
 as functions of temperature for the toy model with the parameters $a = 0.1, b = 0.4, \tilde \Delta=1$.
}
\label{fig_toy_model_1}
\end{minipage}\hfill
\begin{minipage}[t]{.45\linewidth}
\centering
	\includegraphics[width=1\linewidth]{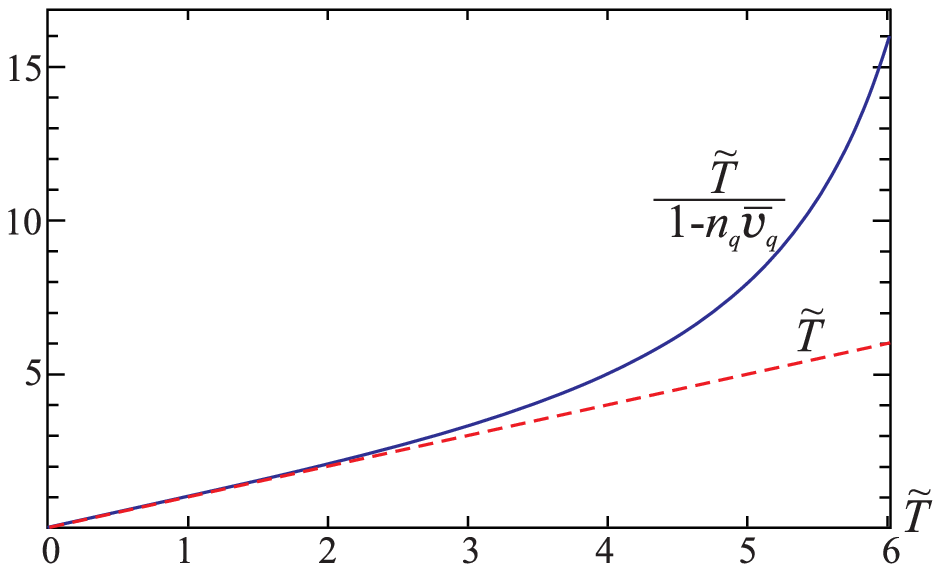}
\caption{
	Comparison of the equations of state of the plasma for the toy model with the internal energy (solid line) and without it (dashed line).
}
\label{eos1}
\end{minipage}
\end{figure}

\begin{figure}[t]
\centering
  \includegraphics[height=5cm]{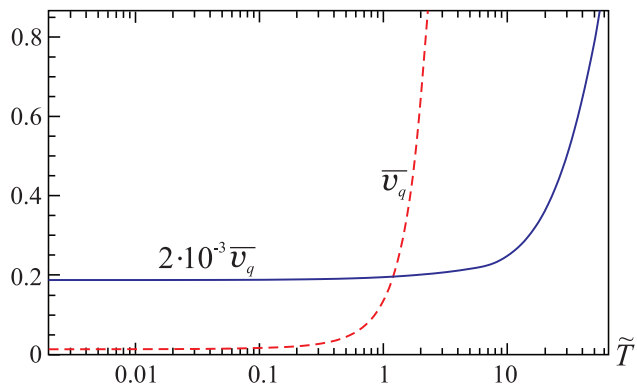}
\caption{
	The average volume $\overline{v_q}$  as a function of temperature for the toy (dashed line) and spinball plasma (solid line) models
	(the spinball plasma model is considered in Sec.~\ref{toy_model_2} below).
}
\label{fig_aver_vol}
\end{figure}

In the above integrals, the integration is taken over the range $-1<x<1$.
The results of numerical calculations for the internal energy, the equation of state, and the average volume of quasiparticles are given in Figs.~\ref{fig_toy_model_1}-\ref{fig_aver_vol}, respectively.
Also, Fig.~\ref{fig_toy_model_1} shows the contribution to the entropy
\begin{equation}
		S = k \ln Z + U/T
\label{entrop}
\end{equation}
coming from the internal energy.

According to the equation of state \eqref{6_a_35} and Fig.~\ref{eos1},
within the toy model under consideration, there is a phase transition when
$n_q \overline{v_q} \rightarrow 1$ and the pressure diverges.
This means that the formula \eqref{6_a_30} for the internal energy can be used only up to the temperature
 $T \lesssim T_c$, where $T_c$ is the critical temperature at which the phase transition occurs.

The main results for the toy model are as follows:
\begin{itemize}
\item
	At zero temperature, the average volume of quasiparticles is not zero, $\overline{l^3(0)} \neq 0$ (see Fig.~\ref{fig_aver_vol}).
Apparently, this is due to the presence of the mass gap in the energy spectrum of the quasiparticles.
\item
	In the presence of the mass gap $\Delta$ the internal energy must be calculated using the formula~\eqref{1_33}. Otherwise,  $U(0) \neq 0$, and when one defines the entropy
using the standard formula~	\eqref{entrop}, the second term in this expression diverges.
\item
	According to Eq.~\eqref{2_110}, there is some temperature  $T_c$, defined from the expression
	\begin{equation}
		\overline{v_q} (T_c) \approx \frac{1}{n_q} ,
	\label{6_a_50}
	\end{equation}
	at which a phase transition takes place. Before the phase transition $T < T_c$ and the spinball plasma consists of
	quasiparticles (spinballs); after the phase transition the particlelike objects (spinballs) are destroyed and the matter has another equation of state.
\end{itemize}

\subsection{Spinball plasma}
\label{toy_model_2}

In this section we consider a more realistic model within which the QGP is modeled by a quark-Proca-gluon plasma, which, in turn, is simplified
so that from all quasiparticles only spinballs remain. By the quark-Proca-gluon plasma we mean
a plasma where quasiparticles are described by particlelike solutions obtained within non-Abelian Proca-Dirac-Higgs theory.

For convenience of performing numerical calculations, we rescale the integrand variables $W_q$ and $T$ in Eq.~\eqref{1_30} as follows:
$
	\bar W_q = \tilde W_q / \tilde \Delta ,
	\bar T = \tilde T /  \tilde \Delta = k T / \Delta
$, where the dimensionless $\tilde \Delta = \Delta/kT_0$ and $\tilde T$ are defined after Eq.~\eqref{6_a_40}
and the dimensionless $\tilde W_q$
is taken from Eq.~\eqref{2_70}. Then the statistical integral $Z_q$ from \eqref{1_30}
takes the form
\begin{equation}
	Z_q \sim \tilde \Delta^{3/2} \int \bar{W}_q^{3/2}(\bm{\gamma})
	e^{- \frac{\bar W_q(\bm{\gamma})}{\bar T}}
	d \bm{\gamma}.
\label{5_b_5}
\end{equation}
Table~\ref{mgsp} shows the values of the energy $\tilde W_q$ for different values of the parameter  $\bar E$. Using these data, one can fit the dimensionless function $\tilde W_q(\bar E)$ in the form
\begin{equation}
	\tilde W_q = \frac{W_0}{\left( 1 - x \right)^\alpha} +
	\frac{W_1}{x^\beta}
\label{5_b_10}
\end{equation}
with the following magnitudes of the fitting parameters:
$	W_0 \approx 2.4,
	W_1 \approx 42.28,
	\alpha \approx 0.6,$ and
	$\beta \approx 2.64$.
Here $x$ corresponds to the parameter $\bar{E}$.

Because of the presence of the internal structure of quasiparticles, the equation of state is modified as follows:
\begin{equation}
	p_q = - \frac{\partial F_q}{\partial V} =
	k T \frac{\partial \ln Z_q}{\partial V} =
	 k N_q \frac{T}{V - V_q},
\label{5_b_30}
\end{equation}
where we have used the expression \eqref{3_15}. Remembering that the volume  $V_q$ can be estimated as
$V_q \approx N_q \overline{ v_q }$,
where $N_q$ is the quasiparticle number and
$\overline{ v_q }(T)$ is the average volume of one quasiparticle that depends on temperature,
we can rewrite Eq.~\eqref{5_b_30} in the form
\begin{equation}
	p_q = \frac{n_q k T}{1 - n_q \overline{v_q } (T) } =
	\frac{\tilde{n}_q  \bar{T}}{1 - \tilde{n}_q  \; \widetilde{\overline{v_q }} (\bar{T}) }
	\frac{\Delta}{l_0^3} \equiv \overline{p}_q\frac{\Delta}{l_0^3},
\label{5_b_50}
\end{equation}
where we have introduced the dimensionless concentration $\tilde{n}_q=l_0^3 n_q$  and average volume
 $\widetilde{\overline{v_q }}=\overline{v_q }/l_0^3 $ and separated out the dimensionless pressure $\overline{p}_q $ and the coefficient which converts the dimensionless quantities into the dimensional ones.
In order to calculate $\overline{ v_q }$ in Eq.~\eqref{5_b_50}, we employ Eq.~\eqref{6_a_40} with the function
\begin{equation}
	\frac{l_q}{l_0}  \approx  \frac{0.94}{ x^{1.09}}   +  \frac{0.4}{\left( 1 - x \right)^{0.58}}
	\label{5_b_60}
\end{equation}
which smoothly interpolates the characteristic sizes $l_1$ and $l_2$ shown in Fig.~\ref{l_1_2}.

\begin{figure}[t]
\begin{minipage}[t]{.45\linewidth}
\centering
	\includegraphics[width=1\linewidth]{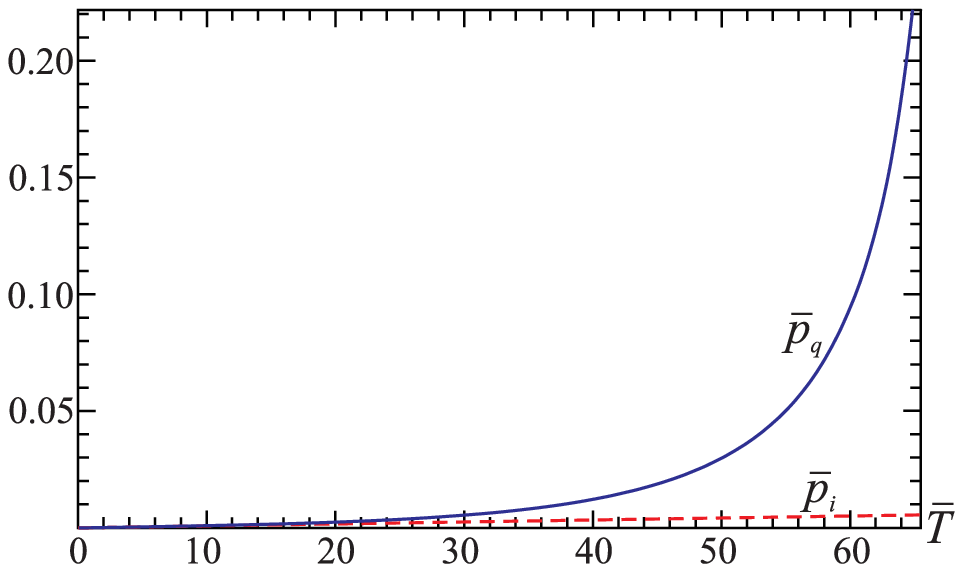}
\caption{
	The solid line shows the equation of state of the spinball plasma \eqref{5_b_50} (when the internal structure is taken into account), the dashed line~-- that of an ideal gas
$\bar{p}_i= \tilde{n}_i  \bar{T}$.
The concentrations $\tilde{n}_q $ and $\tilde{n}_i$ are taken to be equal to $0.85 \cdot 10^{-4}$.
}
\label{eos2}	
\end{minipage}\hfill
\begin{minipage}[t]{.45\linewidth}
\centering
\includegraphics[width=0.98\linewidth]{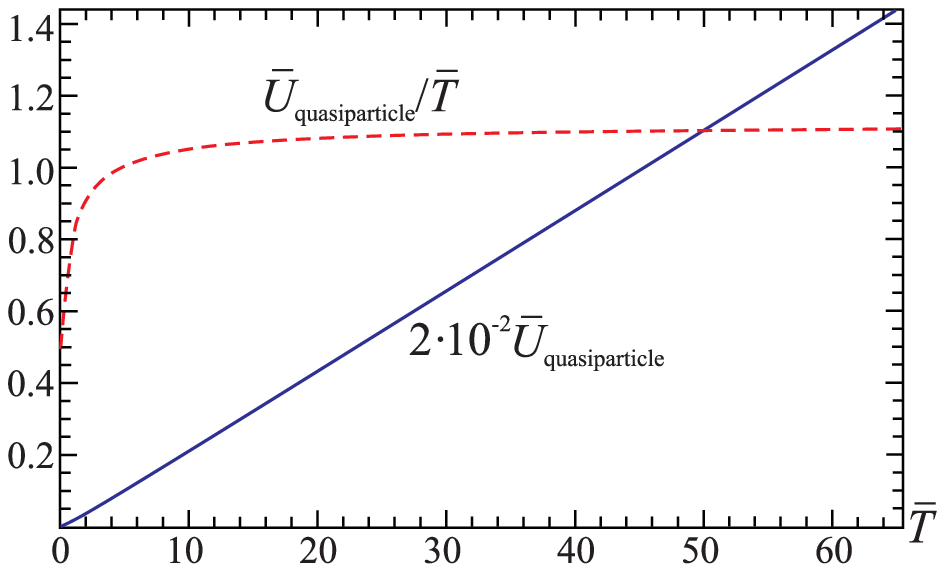}
\caption{
The internal energy $\bar U_{\text{quasiparticle}}\equiv \tilde U_{\text{quasiparticle}}/\tilde{\Delta}$ from \eqref{5_b_55}
and the contribution $\bar U_{\text{quasiparticle}}/\bar T$ to the entropy	\eqref{entrop}
 as functions of temperature for the spinball plasma.
}
\label{fig_toy_model_2}
\end{minipage}
\end{figure}

Notice that from the physical point of view (in the spirit of the van der Waals equation)
the expression
$
 \left( n_q l_0^3 \right)
	\frac{\overline{ v_q }}{l_0^3}
$ in Eq.~\eqref{5_b_50}
is nothing but the fraction of the volume $V$ occupied by the quasiparticles. It is evident that when this fraction tends to unity a phase transition must occur. Fig.~\ref{eos2} shows the pressure $\overline{p}_q$
as compared with the pressure of an ideal gas $\bar{p}_i\equiv p_i/\left(\Delta/l_0^3\right) = \tilde{n}_i  \bar{T}$. For temperatures when 
$
	n_q  \overline{ v_q } (T) \sim 1
$,  the equation of state of the spinball plasma differs considerably from that of an ideal gas.

Let us now estimate dimensional values of the physical parameters of the system under consideration. To do this, we use the expression for the temperature
$T=T_0 \tilde{\Delta} \bar{T}$. Let us suppose that the characteristic temperature $T_0$ is defined by the temperature of quark-gluon plasma
$T_{\text{QGP}}$ (which is obtained from lattice calculations and experiments) as
$T_0=\beta T_{\text{QGP}}$, where $\beta$ is a constant. Then, taking into account that the magnitude of the mass gap
$\tilde{\Delta}\equiv \text{min}\{\tilde{W}_q\}\approx 63$ (this value can be found from interpolation of the data for $\tilde{W}_q$ given
in Table~\ref{mgsp}), for the temperature, say,
$\bar{T}\approx 60$ (see Fig.~\ref{eos2}) one can obtain $T\approx 4\cdot 10^3\beta T_{\text{QGP}}$.
In order to ensure a temperature of our plasma of the order of $T_{\text{QGP}}$, we have to put
$\beta \sim 10^{-3}-10^{-4}$.

Next, let us suppose that the characteristic size $l_0$ is comparable in order of magnitude to a typical QCD length scale
 $l_{\text{QCD}}\sim 1\, \text{fm}$, i.e.,
$l_0=\alpha l_{\text{QCD}}$, where $\alpha $ is some constant of the order of unity. Then
Eq.~\eqref{5_b_50} can be rewritten as follows:
$$
 p_q=\overline{p}_q\frac{\beta}{\alpha^3}\frac{k T_{\text{QGP}}}{ l_{\text{QCD}}^3}\tilde{\Delta}.
$$
For the value of the estimated temperature $\bar{T}\approx 60$ taken above, the dimensionless pressure $\overline{p}_q\approx 0.1$.
Then if we use the values of $ T_{\text{QGP}}\sim 10^{10}-10^{12}\,\text{K}$ obtained in experiments and confirmed by lattice calculations,
then we can find from the above expression that $ p_q\sim 10^{26}-10^{30}\,\text{Pa}$, which agrees well with experimental data.

Finally, according to Eq.~\eqref{1_37}, the contribution to the internal energy from one quasiparticle,
related to the internal structure of the quasiparticle, is defined by the formula
 \begin{equation}
	\tilde U_{\text{quasiparticle}}(\tilde T)\equiv\frac{\tilde U_{\text{quasiparticles}}(\tilde T)}{N_q} = 	
	\frac{
		\int \tilde W_q^{5/2}\left( \bm{\gamma} \right)
		e^{- \frac{\tilde W_q\left( \bm{\gamma} \right)}{\tilde  T}} d \bm{\gamma} 		
	}
	{
		\int \tilde  W_q^{3/2}\left( \bm{\gamma} \right)
		e^{- \frac{\tilde W_q\left( \bm{\gamma} \right)}{\tilde  T}} d \bm{\gamma}
	} -\tilde \Delta,
\label{5_b_55}
\end{equation}
where we have used the dimensionless variables given after Eq.~\eqref{6_a_40}.
The behavior of this term
is shown in Fig.~\ref{fig_toy_model_2}.

It may also be noted that, according to the equation of state \eqref{5_b_50} and Fig.~\ref{eos2},
in the spinball plasma, as in the case of the toy model considered in Sec.~\ref{toy_model_1},
there is a phase transition when $n_q \overline{v_q} \rightarrow 1$
and the pressure $p_q \rightarrow \infty$. This means that the formula \eqref{5_b_55} can be applied to calculate the contribution to the internal energy
only up to the temperature  $T \lesssim T_c$,
and the critical temperature $T_c$ can be defined by the relationship \eqref{6_a_50}.

\section{Conclusions}

In the present study, we have been continuing investigations within the framework of the model of
Ref.~\cite{Dzhunushaliev:2019ham} where a possible relationship between Proca-Dirac-Higgs theory and QCD is suggested.
It is argued in Ref.~\cite{Dzhunushaliev:2019ham} that the presence of a nonlinear spinor field
results in the appearance of a mass gap in  Proca-Dirac-Higgs theory. Also, it is assumed there that the nonlinear Dirac equation
may occur as a result of approximate description of quantum nonperturbative effects in QCD in describing the interaction between
sea quarks and gluons.

On the basis that there may be some correspondence between the quark-gluon plasma in QCD
and the quark-Proca-gluon plasma in non-Abelian Proca-Dirac-Higgs theory, in the present paper,
we have studied the influence that the internal structure of quasiparticles (i.e., their own energy and finite sizes) has on thermodynamics of the QGP.
We have shown that this causes changes both in the internal energy and in the equation of state of the QGP.
In doing so, we take into account the finiteness of sizes of the quasiparticles in the spirit of the van der Waals equation by means of the substraction
of some volume from the total volume occupied by the QGP. But, in contrast to the case of real gases
described by the  van der Waals equation, quasiparticles in the QGP have different sizes depending on the own energy of a given quasiparticle.
It is evident that when the temperature varies the average size of a quasiparticle will also change, resulting
in the corresponding change in the internal energy and pressure of the QGP.

It appears that quasiparticles  (spinballs) in such plasma also play another important role: when one supplies/removes heat, a part of it is used to change
the own energy of the quasiparticles. When the plasma is heated, there is a temperature at which a phase transition takes place.
  This phase transition occurs because
the average size of quasiparticles
increases with temperature, and at some temperature the volume of all quasiparticles begins to approach the volume of the plasma.
As a result, a description of the plasma at a microscopic level, using quasiparticles, becomes inapplicable.
After the phase transition, quasiparticles disappear, and one has to deal with another microscopic description of the QGP.


In summary, the main results of the studies are as follows:
\begin{itemize}
\item
	We use a possible analogy between the QGP in QCD and the plasma in non-Abelian Proca-Dirac-Higgs theory proposed in Ref.~\cite{Dzhunushaliev:2019ham}.
\item
	The expressions for the internal energy and pressure of the QGP containing quasiparticles have been obtained.
\item
	As the simplest approximation for such a plasma, we have considered the case where quasiparticles are spinballs that are described by particlelike solutions
within non-Abelian Proca-Dirac-Higgs theory containing only a nonlinear spinor field. One can assume that such a spinball may describe a pair of sea quarks.
\item
	For the spinball plasma, we have numerically obtained the dependencies of the internal energy and pressure on temperature.
\item
	Within this approximation, it is shown that there is a phase transition from a state with quasiparticles to a state where they are absent.
\end{itemize}

As a final remark, let us note that according to our assumption about a possible relationship between Proca-Dirac-Higgs theory and QCD,
similar processes must take place both in the QGP and in QCD.
This means that the QGP contains quasiparticles  (perhaps they are analogues of Proca monopoles, Proca dyons, spinballs, etc.)
which cause a significant change in thermodynamics of the QGP.
In particular, the presence of the phase transition has the result that a gluon field supporting quasiparticles rearranges into some other structures.
For instance, a transition from a hadronic state to a plasma can perhaps be explained by the fact that in the hadronic state at least some part of
the gluon field is placed inside flux tubes between quarks. When the temperature increases, the cross section of such tubes grows up to some
temperature at which they will occupy almost the entire volume; this leads to the phase transition, after which the gluon field is rearranged into quasiparticles.

\section*{Acknowledgments}
We gratefully acknowledge support provided by Grant No.~BR05236730
in Fundamental Research in Natural Sciences by the Ministry of Education and Science of the Republic of Kazakhstan. V.D. and V.F.
are also grateful to the Research Group Linkage Programme of the Alexander von Humboldt Foundation for the support of this research
and would like to thank the Carl von Ossietzky University of Oldenburg for hospitality while this work was carried out.

\end{document}